\documentclass[sigconf, dvipsnames]{acmart}

\AtBeginDocument{%
  }

\setcopyright{acmlicensed}
\copyrightyear{2025}
\acmYear{2025}
\setcopyright{cc}
\setcctype{by}
\acmDOI{10.1145/3779657.3779670}
\acmConference[WSSE 2025]{2025 The 7th World Symposium on Software Engineering}{October 24--26, 2025}{Okayama, Japan}
\acmBooktitle{2025 The 7th World Symposium on Software Engineering (WSSE 2025), October 24--26, 2025, Okayama, Japan}
\acmISBN{979-8-4007-1577-8/2025/10}

\usepackage{algpseudocode}
\usepackage{algorithm}
\usepackage{comment}
\usepackage[subtle]{savetrees}

\usepackage{listings,jvlisting}
\makeatletter
\let\old@lstKV@SwitchCases\lstKV@SwitchCases
\def\lstKV@SwitchCases#1#2#3{}
\makeatother
\usepackage{lstlinebgrd}
\makeatletter
\let\lstKV@SwitchCases\old@lstKV@SwitchCases

\lst@Key{numbers}{none}{%
    \def\lst@PlaceNumber{\lst@linebgrd}%
    \lstKV@SwitchCases{#1}%
    {none:\\%
     left:\def\lst@PlaceNumber{\llap{\normalfont
                \lst@numberstyle{\thelstnumber}\kern\lst@numbersep}\lst@linebgrd}\\%
     right:\def\lst@PlaceNumber{\rlap{\normalfont
                \kern\linewidth \kern\lst@numbersep
                \lst@numberstyle{\thelstnumber}}\lst@linebgrd}%
    }{\PackageError{Listings}{Numbers #1 unknown}\@ehc}}
\makeatother

\definecolor{lightgray}{rgb}{.9,.9,.9}
\definecolor{darkgray}{rgb}{.4,.4,.4}
\definecolor{purple}{rgb}{0.65, 0.12, 0.82}

\lstdefinelanguage{CSharp}{
  keywords={typeof, new, true, false, catch, return, null, catch, switch, var, if, in, while, do, for, foreach, else, case, break, var, int, float, double, string, public, private, protected, void, static, continue, where, delegate, char},
  keywordstyle=\color{blue}\bfseries,
  ndkeywords={class, export, throw, implements, import, this, ExampleClass, T},
  ndkeywordstyle=\color{darkgray}\bfseries,
  identifierstyle=\color{black},
  sensitive=false,
  comment=[l]{//},
  morecomment=[s]{/*}{*/},
  commentstyle=\color{purple}\ttfamily,
  stringstyle=\color{red}\ttfamily,
  morestring=[b]',
  morestring=[b]"
}

\begin{document}

\title{Code Clone Refactoring in C\# with Lambda Expressions}

\author{Takuto Kawamoto}
\email{tk-kawam@ist.osaka-u.ac.jp}
\orcid{1234-5678-9012}
\author{Yoshiki Higo}
\authornotemark[1]
\email{higo@ist.osaka-u.ac.jp}
\affiliation{%
  \institution{the University of Osaka}
  \city{Suita}
  \state{Osaka}
  \country{Japan}
}

\renewcommand{\shortauthors}{Kawamoto et al.}

\begin{abstract}
``Extract Method'' refactoring is a technique for consolidating code clones. 
Parameterization approaches are used to extract a single method from multiple code clones that contain differences. 
This approach parameterizes expressions and behaviors within a method.
In particular, behavior parameterization has been extensively studied in Java programs, but little research has been conducted on other programming languages.

Lambda expressions can be used to parameterize behaviors, but the specifications of each programming language significantly affect the applicability of this technique. Therefore, the optimal ``Extract Method'' approach may vary depending on the programming language.

In this study, we propose a C\#-specific technique that uses lambda expressions to analyze and consolidate code clones.
We evaluated our proposed method by applying it to code clones detected by the NiCad clone detector and measuring how many of them could be successfully consolidated.

In total, 2,217 clone pairs from 22 projects were included in our evaluation.
For the clone pairs determined to be refactorable, we also attempted refactoring actually.
The proposed approach determined that 35.0\% of all clone pairs were suitable for refactoring.
Among these, 28.9\% were successfully refactored.
\end{abstract}

\keywords{Refactoring, Code clone, Lambda expression}

\maketitle

\section{INTRODUCTION}
Code clones are similar code fragments within the source code.
Pairs of such code fragments are called clone pairs~\cite{baxter}.
Refactoring code clones reduces the size of the code base, helps prevent inconsistent modifications to code clones~\cite{percentage}, and lowers maintenance costs~\cite{cloneEffect}.

``Extract Method'' is a refactoring technique that consolidates code clones into a single method~\cite{extractMethod}.
Since clones often contain minor differences, this technique parameterizes the differing portions to unify the similar fragments~\cite{roy}.

When differences occur at the statement level, expression parameterization may be insufficient to extract them.
This is because expressions are smaller units than statements.
One approach to dealing with statement-level differences is to move the differing statements outside the consolidated code fragment.
However, this reordering can alter program behavior due to side effects, requiring careful analysis to ensure correctness~\cite{type2and3}.
Previous studies have investigated the feasibility of the Extract Method refactoring through statement reordering and expression parameterization, using data flow analysis~\cite{type2and3}.

Some code clones cannot be consolidated simply by reordering statements. 
Alternatively, refactoring approaches that avoid statement reordering may instead parameterize the statements themselves.
Parameterization of statements is called behavior parameterization.

Parameterized expressions are evaluated when the extracted method is invoked, which may alter the original timing of execution~\cite{lambda}.
In contrast, parameterized behavior does not change the execution order.
Lambda expressions are used to parameterize behaviors.
Language-specific features significantly affect how behavior can be parameterized using lambda expressions. Existing research~\cite{lambda} has examined behavior parameterization using lambda expressions in Java. However, it has not addressed its applicability to other programming languages.

This study proposes techniques for behavior parameterization using lambda expressions in C\#. 
Using our proposed method, we analyzed 2,217 clone pairs collected from 22 projects on GitHub to assess their refactorability and performed refactoring on those deemed suitable.

The contributions of this study are as follows:
\begin{itemize}
\item
We provide useful techniques for Extract Method refactoring based on behavior parameterization.
\item
We demonstrate the effectiveness of behavior parameterization in Extract Method refactoring.
Our results suggest that it broadens the range of code clones that are refactorable.
\end{itemize}

This paper is organized as follows: Section 2 introduces background and preparatory concepts for Extract Method refactoring. 
Section 3 presents the proposed approach.
Section 4 describes the evaluation and discusses the results.
Section 5 discusses the results.
Section 6 considers threats to the validity of the study.
Section 7 summarizes the conclusions.

\section{PREPARATION}
\subsection{Code clone}
Code clones are similar code fragments within the source code. Pairs of code fragments that are similar to each other are called clone pairs~\cite{baxter}. Inconsistent changes between code clones can result in missed updates, potentially leading to bugs~\cite{codeClonesAnalysis}.
Based on their degree of similarity, code clones are classified into four types~\cite{roy}.

\begin{enumerate}
\item[Type-1]
The code fragments are entirely identical except for the changes that may exist in whitespace and comments.
\item[Type-2]
The structure of the code fragments is the same, but the identifier names, types, whitespace, and comments may differ.
\item[Type-3]
In addition to changes in identifier names, types, whitespace, and comments, some statements may have been modified or deleted, and others may have been newly added.
\item[Type-4]
The code fragments differ syntactically, but are semantically equivalent.
\end{enumerate}

\subsection{Refactoring code clones}
The ``Extract Method'' refactoring is used to consolidate code clones into a single method.
Parameters can be introduced to apply the ``Extract Method'' refactoring to Type-1 and Type-2 code clones~\cite{type2Refactor}.
The parameters are evaluated when the extracted method is called, which may differ from the timing in the original source code. If the parameterized expressions include method calls or object instantiations, which may have side effects, the difference in evaluation order may affect the program’s behavior~\cite{lambda}.

Prior studies have found that most syntactically similar code clones are Type-2 or Type-3, while Type-1 clones are relatively rare~\cite{type2and3}.
Therefore, refactoring techniques that can consolidate Type-2 and Type-3 code clones play an important role.

\subsection{``Extract Method'' with behavior parameterization}
Some code clones cannot be consolidated into a single method, even with expression parameterization.
Figure~\ref{lst:example-case} shows such an example.
In these code fragments, the highlighted parts differ, which classifies them as Type-3 code clones.

These code fragments cannot be consolidated into a single method with expression parameterization. 
They cannot be parameterized because the differing parts are statements rather than expressions.
A combined behavior and expression parameterization approach has been proposed~\cite{lambda}.
Function objects, such as lambda expressions, defer evaluation until runtime, thereby preserving program behavior.

In some programming languages, lambda expressions can capture variables and functions from the scope in which the code clone appears.
For example, lambda expressions can be used for behavior parameterization in Java, C\#, and Python.
In JavaScript, arrow functions serve the same purpose.
For simplicity, we collectively refer to such function objects, including arrow functions, as lambda expressions throughout this study.

Figure~\ref{lst:example-refactoring} shows an example of method refactoring using behavior parameterization, based on the code in Figure~\ref{lst:example-case}.
The common parts between the code fragments are highlighted in orange. Furthermore, the differing parts are highlighted in distinct colors.
In both code fragments, the original code has been rewritten to call the newly generated method \texttt{Subroutine}, where the second argument is a lambda expression representing the differing parts of the original source code.
The definition of \texttt{Subroutine} is shown in Figure~\ref{lst:example-method}. 
The definition contains the common parts of both code fragments and invokes to the lambda expressions passed as arguments to replace the differing parts.

The lambda expression may be executed multiple times during a single run of the code fragment, outputting a string to the console each time.
This behavior cannot be achieved with simple expression parameterization.

Previous research~\cite{lambda} proposed a technique for ``Extract Method'' refactoring using lambda expressions in Java. However, this technique cannot be directly applied to code clones in other programming languages.
This is because the proposed technique is specific to the Java language.
For example, this technique accounts for checked exceptions and final variables~\cite{javaLambdaExp}, both of which are not present in C\#.
Instead, C\# includes features such as properties, which require careful consideration when performing expression parameterization.
According to a recent review~\cite{review}, apart from the technique proposed for Java, no other refactoring techniques using lambda expressions have been proposed.
Therefore, this study proposes a technique for ``Extract Method'' refactoring using lambda expressions in C\#. 
Additionally, we examine the applicability of the proposed technique to other programming languages.

For simplicity, we refer to ``Extract Method'' refactoring using behavior parameterization simply as ``Extract Method'' refactoring, and to the extracted method as a subroutine.

\begin{figure}[tb]
\begin{lstlisting}[language=CSharp,linebackgroundcolor={%
\ifnum\value{lstnumber}=1\color{BurntOrange}\fi%
\ifnum\value{lstnumber}=2\color{BurntOrange}\fi%
\ifnum\value{lstnumber}=3\color{BurntOrange}\fi%
\ifnum\value{lstnumber}=4\color{SeaGreen}\fi%
\ifnum\value{lstnumber}=5\color{BurntOrange}\fi%
\ifnum\value{lstnumber}=6\color{BurntOrange}\fi%
}]
StreamReader reader = new StreamReader("Path");
while (reader.Peek() != -1) {
    string line = reader.ReadLine();
    Console.WriteLine(line);
}
reader.Close();
\end{lstlisting}
\begin{lstlisting}[language=CSharp,linebackgroundcolor={%
\ifnum\value{lstnumber}=1\color{BurntOrange}\fi%
\ifnum\value{lstnumber}=2\color{BurntOrange}\fi%
\ifnum\value{lstnumber}=3\color{BurntOrange}\fi%
\ifnum\value{lstnumber}=4\color{Turquoise}\fi%
\ifnum\value{lstnumber}=5\color{Turquoise}\fi%
\ifnum\value{lstnumber}=6\color{Turquoise}\fi%
\ifnum\value{lstnumber}=7\color{BurntOrange}\fi%
\ifnum\value{lstnumber}=8\color{BurntOrange}\fi%
}]
StreamReader reader = new StreamReader("OtherPath");
while (reader.Peek() != -1) {
    string line = reader.ReadLine();
    for (int i = 0;i < line.Length; i++) {
        Console.WriteLine(line[i]);
    }
}
reader.Close();
\end{lstlisting}
\caption{Examples difficult to extract via expression parameterization}
\label{lst:example-case}
\end{figure}

\begin{figure}[tb]
\begin{lstlisting}[language=CSharp,linebackgroundcolor={%
\ifnum\value{lstnumber}=2\color{SeaGreen}\fi%
}]
Subroutine("Path", (line) => {
    Console.WriteLine(line);
});
\end{lstlisting}
\begin{lstlisting}[language=CSharp,linebackgroundcolor={%
\ifnum\value{lstnumber}=2\color{Turquoise}\fi%
\ifnum\value{lstnumber}=3\color{Turquoise}\fi%
\ifnum\value{lstnumber}=4\color{Turquoise}\fi%
}]
Subroutine("OtherPath", (line) => {
    for (int i = 0;i < line.Length; i++) {
        Console.WriteLine(line[i]);
    }
});
\end{lstlisting}
\caption{Example of refactoring Figure 1}
\label{lst:example-refactoring}
\end{figure}

\begin{figure}
\begin{lstlisting}[language=CSharp,linebackgroundcolor={%
\ifnum\value{lstnumber}=3\color{BurntOrange}\fi%
\ifnum\value{lstnumber}=4\color{BurntOrange}\fi%
\ifnum\value{lstnumber}=5\color{BurntOrange}\fi%
\ifnum\value{lstnumber}=7\color{BurntOrange}\fi%
\ifnum\value{lstnumber}=8\color{BurntOrange}\fi%
}]
delegate void LambdaType(string v);
void Subroutine(string v1, LambdaType lambda1) {
    StreamReader reader = new StreamReader(v1);
    while (reader.Peek() != -1) {
        string line = reader.ReadLine();
        lambda1.Invoke(line);
    }
    reader.Close();
}
\end{lstlisting}
\caption{Method consolidated with Figure 2}
\label{lst:example-method}
\end{figure}

\section{OUR APPROACH}

\subsection{Overview}
Our approach takes the source code and a Type-1, Type-2, or Type-3 clone pair as input, and applies ``Extract Method'' refactoring to produce the refactored source code.
The code fragments in the clone pair are limited to sequences of one or more consecutive statements.

\begin{figure*}[tb]
  \centering
  \includegraphics[width=1.6\columnwidth]{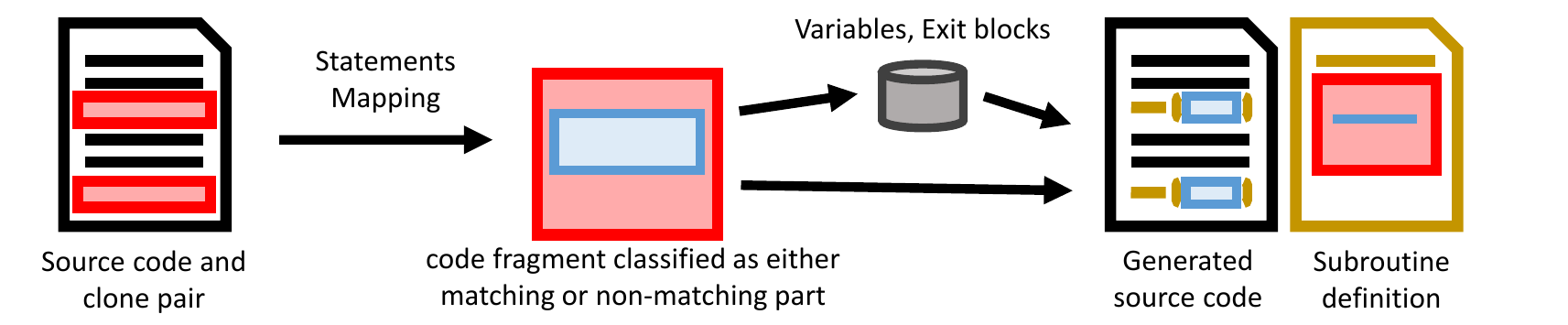}
  \caption{Overview of our approach}
  \label{fig:method}
\end{figure*}

Figure~\ref{fig:method} provides an overview of our approach.
The approach identifies the portions of the given clone pair that should be parameterized using lambda expressions and then generates a subroutine.
Specifically, the subroutine generation process is divided into five steps:
\begin{enumerate}
\item Statement mapping,
\item Variable mapping,
\item Exit block checking,
\item Argument determination, and
\item Source code generation.
\end{enumerate}
In the remainder of this section, we explain each step in detail.

\subsection{Statement mapping}
First, this approach maps the statements in the input clone pair. 
It scans the code fragments from the beginning and identifies matching statement pairs, while ignoring differences in identifiers.

Through this mapping, the statements in each fragment are classified as either matching or non-matching with those in the other fragment.
During code generation, non-matching statements are parameterized using lambda expressions, while matching ones are directly included in the subroutine.

We apply the Longest Common Subsequence (LCS) algorithm to perform this mapping.
In our approach, the LCS algorithm is used to obtain the longest common subsequence of statements, rather than substrings.

Typically, the LCS algorithm is applied to strings to find the longest common subsequence of characters.
It also requires a comparison function for the sequence elements.
When applied to strings, these elements are characters, and thus a simple character-level comparison suffices. 
However, in our case, the elements are program statements, which require a comparison function capable of evaluating their similarity.
The following describes the statement comparison function used in our approach.

\subsubsection{Statement comparator}
The LCS algorithm requires a comparison function to determine equivalence.
We consider statements equivalent if they are so similar that they do not require behavior parameterization.
Our approach uses a comparator that considers statements to match if they differ only in whitespace, type names, or identifiers.

Ultimately, statement comparison is performed using a simple string comparison.
To allow for specific differences, the statements to be compared are normalized. In other words, the comparison function normalizes both statements and then simply compares the normalized versions.

Whitespace and comments can easily be ignored in the abstract syntax tree obtained from the source code.
Furthermore, for identifier normalization, our approach masks identifier names.
For example, constants and variables are masked because they can often be parameterized.

There are constraints on expressions that can be parameterized, so identifier normalization must be handled carefully.
Expressions in the source code are further categorized into finer-grained types, and some grammar rules restrict which types can be used in specific contexts.
For example, expressions used in \texttt{case} labels of \texttt{switch} statements must be constant expressions whose values are determined at compile time.
Since the parameters are clearly not constant expressions, the identifier names appearing within the switch \texttt{case} labels must not be masked.
We conducted a detailed analysis of the grammar rules of C\# and restricted identifier parameterization to those that meet the following criteria:
\begin{enumerate}
\item Member identifiers,
\item Type identifiers,
\item Method and constructor identifiers,
\item Identifiers in \texttt{case} labels of \texttt{switch} statements, and
\item Identifiers used in pattern matching.
\end{enumerate}

\subsubsection{Nested statements matching}
For statements containing multiple inner statements---such as \texttt{for} or \texttt{while} loops---simply determining whether they match is not sufficient.
We refer to the statements that contain inner statements as nested statements.

Partial behavior parameterization may be desirable within a \texttt{for} loop.
For example, most of the inner statements of a \texttt{for} loop may match, while some inner statements may differ.

For nested statements, the comparison is first performed by ignoring their inner statements.
The LCS algorithm is then applied to the inner statement sequences of the nested statements that are determined to match.
This process of applying the LCS algorithm is repeated recursively according to the nesting depth.

The way inner statements are ignored varies depending on the type of statement.
For example, in the case of an \texttt{if} statement, only the conditional expression is compared.
For a \texttt{for} statement, the statement comparator ignores inner statements and considers only the loop variable declaration, the loop continuation condition, and the update statement.

\subsection{Variable mapping}
Second, in expression parameterization, we map corresponding expressions within matching statements. 
These expressions may include local variables, members, fields, and literals.
However, in this study, we exclude expressions that contain operators.

From the matching statements identified in the statement mapping step, we extract parameterizable expressions that occur at the same positions in both statements.
Our approach then assigns parameters to these expression pairs.
Identical expression pairs share the same parameter, while distinct pairs receive different parameters.
Variables that can be reassigned, such as the left-hand side of an assignment statement, should be parameterized by reference.

If a parameter is passed to a subroutine by value, any changes made to it inside the subroutine are not preserved after it exits.
Parameters that are assigned values within the subroutine should be passed by reference.

\subsubsection{Parameter typing}
Each parameter must have a type.
If both expressions being parameterized share the same type, the parameter should also use this type.
If the expressions have different types, we check whether there is a common parent type that can safely be used for both.
If such a type exists, the parameter type is set to this common parent type.
Otherwise, our approach considers using generics as an alternative.

A simple example of a code clone requiring generics is shown in Figure~\ref{lst:generics-case-orig}. The \texttt{Insert} and \texttt{RemoveAt} methods in the code can be used only with lists, whereas the \texttt{Overlaps} method is only applicable to sets. The \texttt{Add} method is used with both lists and sets, and is defined in the common parent interface \texttt{ICollection<int>}.

In the subroutine definition shown in Figure~\ref{lst:generics-method-case}, only methods compatible with \texttt{ICollection<int>} are used.
In contrast, the caller in Figure~\ref{lst:generics-case-refactored} uses methods defined for their respective, more specific types.
Generics are useful when different types need to be used at each call site.
To enable the use of methods defined in a common parent class within the subroutine, the generic variable must be constrained to inherit the required type.

If no suitable common parent type exists and generics cannot be used, an error is raised.
In this case, the refactoring is terminated.

\begin{figure}[tb]
\begin{lstlisting}[language=CSharp,linebackgroundcolor={%
\ifnum\value{lstnumber}=2\color{BurntOrange}\fi%
\ifnum\value{lstnumber}=3\color{SeaGreen}\fi%
\ifnum\value{lstnumber}=4\color{SeaGreen}\fi%
\ifnum\value{lstnumber}=5\color{BurntOrange}\fi%
}]
List<int> list = [1, 2, 3];
list.Add(5);
list.Insert(2, 6);
list.RemoveAt(0);
list.Add(10);
\end{lstlisting}
\begin{lstlisting}[language=CSharp,linebackgroundcolor={%
\ifnum\value{lstnumber}=2\color{BurntOrange}\fi%
\ifnum\value{lstnumber}=3\color{Turquoise}\fi%
\ifnum\value{lstnumber}=4\color{BurntOrange}\fi%
}]
HashSet<int> set = [4, 3, 6];
set.Add(5);
set.Overlaps([1, 2, 3]);
set.Add(10);
\end{lstlisting}
\caption{Code clones requiring generics}
\label{lst:generics-case-orig}
\end{figure}

\begin{figure}[tb]
\begin{lstlisting}[language=CSharp,linebackgroundcolor={%
\ifnum\value{lstnumber}=3\color{SeaGreen}\fi%
\ifnum\value{lstnumber}=4\color{SeaGreen}\fi%
}]
List<int> list = [1, 2, 3];
Subroutine<List<int>>(list, (c) => {
    c.Insert(2, 6);
    c.RemoveAt(0);
});
\end{lstlisting}
\begin{lstlisting}[language=CSharp,linebackgroundcolor={%
\ifnum\value{lstnumber}=3\color{Turquoise}\fi%
}]
HashSet<int> set = [4, 3, 6];
Subroutine<HashSet<int>>(set, (c) => {
    c.Overlaps([1, 2, 3]);
});
\end{lstlisting}
\caption{Example of refactoring Figure 5}
\label{lst:generics-case-refactored}
\end{figure}

\begin{figure}[tb]
\begin{lstlisting}[language=CSharp,linebackgroundcolor={%
\ifnum\value{lstnumber}=3\color{BurntOrange}\fi%
\ifnum\value{lstnumber}=5\color{BurntOrange}\fi%
}]
delegate void LambdaType<T>(T v);
void Subroutine<T>(T collection, LambdaType<T> lambda) where T : ICollection<int> {
    collection.Add(5);
    lambda.Invoke(collection);
    collection.Add(10);
}
\end{lstlisting}
\caption{Method consolidated with Figure 6}
\label{lst:generics-method-case}
\end{figure}

\begin{figure}[tb]
\begin{lstlisting}[language=CSharp,linebackgroundcolor={%
\ifnum\value{lstnumber}=2\color{BurntOrange}\fi%
\ifnum\value{lstnumber}=3\color{BurntOrange}\fi%
\ifnum\value{lstnumber}=4\color{BurntOrange}\fi%
\ifnum\value{lstnumber}=5\color{BurntOrange}\fi%
\ifnum\value{lstnumber}=6\color{SeaGreen}\fi%
\ifnum\value{lstnumber}=7\color{SeaGreen}\fi%
\ifnum\value{lstnumber}=8\color{BurntOrange}\fi%
\ifnum\value{lstnumber}=9\color{BurntOrange}\fi%
}]
int total = 0;
StreamReader reader = new StreamReader(filePath);
string line;
while ((line = reader.ReadLine()) != null) {
    if (int.TryParse(line, out int number)) {
        if (number == 0) break;
        total += number;
    }
}
\end{lstlisting}
\begin{lstlisting}[language=CSharp,linebackgroundcolor={%
\ifnum\value{lstnumber}=2\color{BurntOrange}\fi%
\ifnum\value{lstnumber}=3\color{BurntOrange}\fi%
\ifnum\value{lstnumber}=4\color{BurntOrange}\fi%
\ifnum\value{lstnumber}=5\color{BurntOrange}\fi%
\ifnum\value{lstnumber}=6\color{Turquoise}\fi%
\ifnum\value{lstnumber}=7\color{Turquoise}\fi%
\ifnum\value{lstnumber}=8\color{BurntOrange}\fi%
\ifnum\value{lstnumber}=9\color{BurntOrange}\fi%
}]
int max = 0;
StreamReader reader = new StreamReader(filePath);
string line;
while ((line = reader.ReadLine()) != null) {
    if (int.TryParse(line, out int number)) {
        if (number < 0) break;
        if (max < number) max = number;
    }
}
\end{lstlisting}
\caption{Code clones containing jump statements}
\label{lst:break-case-orig}
\end{figure}

\begin{figure}[tb]
\begin{lstlisting}[language=CSharp,linebackgroundcolor={%
\ifnum\value{lstnumber}=3\color{SeaGreen}\fi%
\ifnum\value{lstnumber}=4\color{SeaGreen}\fi%
\ifnum\value{lstnumber}=5\color{SeaGreen}\fi%
}]
int total = 0;
Subroutine((number) => {
    if (number == 0) return true;
    total += number;
    return false;
});
\end{lstlisting}
\begin{lstlisting}[language=CSharp,linebackgroundcolor={%
\ifnum\value{lstnumber}=3\color{Turquoise}\fi%
\ifnum\value{lstnumber}=4\color{Turquoise}\fi%
\ifnum\value{lstnumber}=5\color{Turquoise}\fi%
}]
int max = 0;
Subroutine((number) => {
    if (number < 0) return true;
    if (max < number) max = number;
    return false;
});
\end{lstlisting}
\caption{Example of refactoring Figure 8}
\label{lst:break-case-refactored}
\end{figure}

\begin{figure}[tb]
\begin{lstlisting}[language=CSharp,linebackgroundcolor={%
\ifnum\value{lstnumber}=3\color{BurntOrange}\fi%
\ifnum\value{lstnumber}=4\color{BurntOrange}\fi%
\ifnum\value{lstnumber}=5\color{BurntOrange}\fi%
\ifnum\value{lstnumber}=6\color{BurntOrange}\fi%
\ifnum\value{lstnumber}=8\color{BurntOrange}\fi%
\ifnum\value{lstnumber}=9\color{BurntOrange}\fi%
}]
delegate bool LambdaType(int number);
void Subroutine(string path, LambdaType lambda) {
    StreamReader reader = new StreamReader(filePath);        
    string line;
    while ((line = reader.ReadLine()) != null) {
        if (int.TryParse(line, out int number)) {
            if (lambda.Invoke(number)) break;
        }
    }
}
\end{lstlisting}
\caption{Method consolidated with Figure 9}
\label{lst:break-method-case}
\end{figure}

\subsection{Exit block checking}
Handling statements that alter control flow, such as \texttt{break} statements, poses a significant challenge for the Extract Method refactoring.
In a Control Flow Graph, which represents program execution paths as a directed graph, the final node to be executed is called an exit block. 
This study defines an exit block as either the last executable statement in a code fragment or the end of the code fragment if control can reach it. In C\#, such last executable statements include \texttt{break}, \texttt{continue}, \texttt{return}, and \texttt{goto} statements. Different exit blocks lead to distinct subsequent behaviors after exiting the code fragment. Crucially, a code fragment with a single exit block cannot emulate the behavior of one with multiple exit blocks.

A function call statement has only a single exit block, whereas a code fragment composed of several statements may contain multiple exit blocks. The Extract Method refactoring replaces a code fragment with a function call. 
A code fragment with multiple exit blocks cannot be replaced by a simple function call.
Therefore, we emulate the behavior of multiple exit blocks by carefully designing the subroutine's return value and its handling at the call site.

For simplicity, \texttt{break}, \texttt{continue}, and \texttt{return} statements are considered the end of code fragments in this study.
We excluded the \texttt{goto} statement from our analysis. Since \texttt{goto} statements can jump to arbitrary locations and make control flow difficult to track, we omitted them. A more detailed discussion on the impact of including \texttt{goto} is discussed later. This execlusion is justified because the use of \texttt{goto} is widely considered poor practice that reduces code readability~\cite{goto}.

Specifically, the subroutine's return value indicates the type of exit block to the call site. The call site of the subroutine inserts an \texttt{if} statement after the function call to execute the appropriate behavior for each exit block, thereby emulating the original control flow containing multiple exit blocks. 
To ensure that only necessary exit blocks are emulated, it is essential to check the exit blocks included in the code fragments of the clone pair and in parameterized behavior.

Figure~\ref{lst:break-case-orig} illustrates an example requiring the emulation of exit block behavior. The matching code segments are highlighted in orange, while the behaviors parameterized by a lambda expression are highlighted in blue and green. This parameterized behavior contains a break statement, resulting in multiple exit blocks. Figure~\ref{lst:break-case-refactored} shows the source code after refactoring, and Figure~\ref{lst:break-method-case} presents the definition of the extracted subroutine. Within the parameterized statement, the break statement is replaced by a return statement that returns a Boolean value. This return value informs the call site which behavior to emulate.

The extracted method is similar to the common higher-order function \texttt{reduce}. However, unlike \texttt{reduce}, this method can terminate the loop prematurely. Our approach provides developers with insights into refactoring using higher-order functions.

This approach can be a bit tricky to apply in some programming languages.
One notable programming language is Java.
Java supports labeled continue and break statements, which can result in multiple possible target locations for control transfer.
To accurately emulate the behavior of continue and break statements, it is necessary to take labels into account.
JavaScript also has similar characteristics.
A similar problem occurs with the \texttt{goto} statement, which, like labeled statements, enables transfers to multiple destinations. This is why our research does not consider the \texttt{goto} statement.

\subsection{Argument determination}
Arguments must be appropriately passed to both subroutines and lambda expressions used for behavior parameterization.

If a local variable declared in a subroutine is accessed within a parameterized lambda expression, it must be passed as an argument from the subroutine to the lambda expression.
Additionally, the subroutine must receive the lambda expression to be invoked as an argument.

When a lambda expression is passed as an argument for behavior parameterization, its type must also be explicitly specified.
In C\#, delegate types, which are used to reference functions, must be declared in advance.

\subsection{Source code generation}
The subroutine is generated using one of the code fragments in the clone pair as a template.
In our approach, the static class defining the subroutine is placed in the global namespace.
To define this static class, a new source file is added to the project.

In addition to the static class, this source file defines the delegate type for lambda expressions and an enumeration representing exit block types.
Both types are declared with the \texttt{internal} access modifier.
The \texttt{internal} modifier allows access only within the same project.
Compared to the \texttt{public} modifier, \texttt{internal} supports more refactoring scenarios because it avoids the need to expose refactored components to external projects.

\subsection{Related characteristics of C\#}
Our approach relies significantly on the specifications of the programming language.
This section outlines the C\# language features most pertinent to our approach.

\subsubsection{Parameter modifier}
In C\#, parameters are generally passed by value. 
In this case, a function cannot reassign a value to the caller’s variable through a parameter. 
However, by adding the \texttt{ref} modifier to a parameter, the function can pass it by reference, allowing it to modify the caller’s variable.

The ref modifier can only be applied to variables that have already been initialized.
If a variable is uninitialized, the \texttt{out} modifier can be used instead of \texttt{ref}, enabling the function to initialize the variable within its body. 
Our approach uses appropriate parameter modifiers to create subroutines that reassign parameter values without significantly changing the original code structure.

\subsubsection{Restriction of parameter modifier}
C\# supports asynchronous methods and iterator methods, also known as generators in other languages.
However, parameter modifiers cannot be used in these methods.\footnote{As of the latest C\# version released in 2025, parameter modifiers are allowed in asynchronous and iterator methods under certain conditions.}
Therefore, this study excludes from refactoring any code clones that appear in contexts where parameter modifiers cannot be used.

\subsubsection{Property}
A property is a class or structure member consisting of a getter, a setter, or both. Since properties are accessed the same way as fields or variables, it is impossible to distinguish between property access and field or variable access based solely on static syntactic analysis. 
However, unlike fields and other members, properties cannot be passed by reference to functions using parameter modifiers. 
Nevertheless, to achieve pass-by-reference behavior, getter and setter functions must be passed as delegates.

\subsection{Integration with code clone detector}
The proposed approach takes a clone pair as input.
The suitability of the clone pair for Extract Method refactoring is not verified beforehand.
The proposed approach must determine whether the given clone pair is suitable for refactoring, either before or during the process.
During the process, the approach may terminate with an error.
Such an error can be interpreted as an indication that the clone pair is unsuitable for refactoring.

The reasons why a clone pair may be deemed unsuitable are as follows:
\begin{enumerate}
\item Cross-project code clones,
\item No matching statements,
\item Variable type mismatch,
\item The required \texttt{ref} or \texttt{out} modifiers cannot be added,
\item Code clones containing method calls,
\item Code clones containing \texttt{yield}, \texttt{async}, \texttt{await}, and
\item Source code analysis failed.
\end{enumerate}

(1) indicates that no matching statements could be found during statement mapping.
(2) indicates that the parameter types could not be determined during variable mapping.
(3) indicates that \texttt{ref} or \texttt{out} modifiers are required in variable mapping but cannot be applied.
Sometimes, \texttt{ref} and \texttt{out} modifiers cannot be used inside lambda expressions.
(4) indicates that the code clone contains an instance method call.
Since the subroutine is defined in a separate class, the instance methods from the original class cannot be invoked directly.
(5) indicates that the code clones are located within iterator methods or asynchronous methods.
This study excludes code clones within iterator or asynchronous methods from refactoring targets.
(6) represents other abnormal cases, such as broken projects that could not be correctly analyzed.

In addition to performing refactoring, the proposed approach may report one of the errors listed in (1) through (6).
In particular, detecting errors (1) through (5) indicates that the clone pair is either out of scope or not suitable for refactoring.

\section{EVALUATION}
\subsection{Procedure}
To evaluate the accuracy and applicability of the proposed subroutine generator, we applied it to clone pairs extracted from real-world software projects.
We assessed the correctness of the generated subroutines by comparing project test results before and after applying the subroutine generator to the input clone pairs.
If the test results were identical, we determined that the project's behavior was preserved after subroutine generation.

The input code clones for the subroutine generator were obtained from the NiCad code clone detector~\cite{NiCad}, which detects Type-1, Type-2, and Type-3 code clones.
We collected 22 C\# projects from GitHub that include tests and evaluated 2,217 clone pairs detected by NiCad.

However, this simple comparison is not sufficient. If the refactored portions of code are not executed by any test, the behavior may appear unchanged even when the refactoring is incorrect.
To address this, we intentionally applied incorrect refactoring and compared the resulting test outcomes.
If the test results before and after applying the incorrect refactoring differed, we determined that the affected code was exercised by the tests.
From this point onward, we refer to the subroutine generator that applies incorrect refactoring as the incorrect subroutine generator.
The only difference between the subroutine generator and the incorrect subroutine generator is the function body of the generated subroutines.
Specifically, the incorrect subroutine generator generates subroutines that throw exceptions. Both generators output the same source code, except for the function body.
The evaluation procedure using the incorrect subroutine generator is described below.
The authors prepared three versions of the project: 
\begin{enumerate}
\item original project, 
\item project after applying the subroutine generator, and 
\item project after applying the incorrect subroutine generator. 
\end{enumerate}
We ran tests for each version and compared their test results.
If the test results of the original project differed from those of the project with the incorrect subroutine generator, we assumed that the refactored code had been run by the tests.
If those results matched the ones from the project with the correct subroutine generator, we concluded that the subroutine generator correctly preserved the program’s behavior.
To eliminate non-deterministic test results, we ran each test four times and confirmed no variations in the test outcomes. 

\subsection{Results}
Table~\ref{tab:classification} shows the classification of code clones produced by our proposed approach.
Items (1) to (7) correspond to the error categories described in Section 3.8.
35.0\% of the total, 775 to be exact, clone pairs were determined to be refactorable, and were actually refactored.
They were then evaluated using the procedure described earlier.

The evaluation results for each clone pair are shown in Table~\ref{tab:results}.
The clone pairs were classified into three categories based on build success and test results.
“Refactored successfully” refers to clone pairs for which the refactored source code built successfully.
The correctness of the refactoring was also verified through testing.
“Refactored, not tested” refers to clone pairs where the refactored source code built successfully.
However, there were no tests covering the refactored portions, so the correctness of the refactoring could not be confirmed.
“Failed to refactor” refers to clone pairs for which refactoring failed.

Among the refactorable clone pairs, 32.1\% had refactored source code that both built successfully and passed the tests.
In addition, 61.0\% of the refactorable clone pairs built successfully.

\begin{table}[b]
\centering
\caption{Classification Results}
\label{tab:classification}
\begin{tabular}{lr}\hline
Category                  & Code clones \\ \hline
Refactorable code clones         & 775 \\
(1) No matching statements       & 842 \\
(2) Variable type mismatch       & 46 \\
(3) \texttt{ref}, \texttt{out} modifiers cannot be added    & 100 \\
(4) Code clones containing method calls & 119 \\
(5) Code clones containing \texttt{yield}, \texttt{async}, \texttt{await}    & 293 \\
(6) Source code analysis failed    & 42 \\\hline
Total                     & 2,217         \\\hline 
\end{tabular}
\end{table}

\begin{table}[b]
\centering
\caption{Evaluation Results}
\label{tab:results}
\begin{tabular}{lr}\hline
Category                  & Code clones \\ \hline
Refactored successfully   & 249           \\
Refactored, not tested    & 224          \\
Failed to refactor        & 302          \\\hline
Total                     & 775         \\\hline 
\end{tabular}
\end{table}

\section{DISCUSSION}
\subsection{Why do code clone statements not match?}
Table~\ref{tab:classification} shows that in 38.0\% of the clone pairs, no matching statements were found.
Since clone pairs consist of similar groups of statements, the absence of matching statements may seem surprising.
NiCad detects Type-2 code clones, which are identified by ignoring differences in identifier names.
NiCad disregards identifier name differences regardless of the type of identifier.
In contrast, our approach selectively ignores identifier name differences based on the type of identifier.

In practice, some clone pairs included \texttt{switch} statements with differing \texttt{case} labels.
Such differences can explain the absence of matching statements.

Additionally, we observed cases where matches inside nested constructs were missed.
In this study, \texttt{for} statements were treated as one of the constructs that form nested structures. A \texttt{for} statement was considered a match only when its loop variable definition and continuation condition were identical.
If the code fragment consisted solely of \texttt{for} statements and there were slight differences in their continuation conditions, no match was detected.

An example where refactoring was not possible due to minor differences in the continuation conditions of \texttt{for} statements is shown in Figure~\ref{lst:nested-for}.
In this example, all statements inside the for loops match except for differences in identifier names.
However, the loop variable definitions and continuation conditions differ significantly, making it difficult to consolidate the code fragments into a single method through expression parameterization.
Although only the highlighted line containing the \texttt{for} loop condition differs, the rest of the source code is nearly identical.
Nonetheless, the entire code fragment cannot be directly extracted as a single method.
While extracting the entire code fragment as a method is not straightforward, it would be possible to extract only the inner part of the \texttt{for} loop.

As a possible countermeasure, we could consider applying the Extract Method refactoring within the nested structure itself.
This is applicable when the surrounding nested constructs do not match.
However, attempting to refactor every possible combination of nested structures within a code fragment results in exponential growth in computational complexity. 
This complexity increases with the number of nested constructs.

Another possible improvement is to implement more flexible expression parameterization to absorb larger differences in loop variable declarations and continuation conditions.

\begin{figure}[tb]
\begin{lstlisting}[language=CSharp,linebackgroundcolor={%
\ifnum\value{lstnumber}=1\color{SeaGreen}\fi%
\ifnum\value{lstnumber}=2\color{BurntOrange}\fi%
\ifnum\value{lstnumber}=3\color{BurntOrange}\fi%
\ifnum\value{lstnumber}=4\color{BurntOrange}\fi%
\ifnum\value{lstnumber}=5\color{BurntOrange}\fi%
\ifnum\value{lstnumber}=6\color{BurntOrange}\fi%
\ifnum\value{lstnumber}=7\color{BurntOrange}\fi%
\ifnum\value{lstnumber}=8\color{BurntOrange}\fi%
\ifnum\value{lstnumber}=9\color{BurntOrange}\fi%
\ifnum\value{lstnumber}=10\color{BurntOrange}\fi%
\ifnum\value{lstnumber}=11\color{BurntOrange}\fi%
\ifnum\value{lstnumber}=12\color{SeaGreen}\fi%
}]
for (int i = 0; i < str.Length; i++) {
    char c = str [i];
    if (c == '\n') {
        if (idx == -1)
            idx = i;
        idx = i;
    }
    else if (c == ' ' || c == '\t')
        continue;
    else
        break;
}
\end{lstlisting}
\begin{lstlisting}[language=CSharp,linebackgroundcolor={%
\ifnum\value{lstnumber}=1\color{Turquoise}\fi%
\ifnum\value{lstnumber}=2\color{BurntOrange}\fi%
\ifnum\value{lstnumber}=3\color{BurntOrange}\fi%
\ifnum\value{lstnumber}=4\color{BurntOrange}\fi%
\ifnum\value{lstnumber}=5\color{BurntOrange}\fi%
\ifnum\value{lstnumber}=6\color{BurntOrange}\fi%
\ifnum\value{lstnumber}=7\color{BurntOrange}\fi%
\ifnum\value{lstnumber}=8\color{BurntOrange}\fi%
\ifnum\value{lstnumber}=9\color{BurntOrange}\fi%
\ifnum\value{lstnumber}=10\color{BurntOrange}\fi%
\ifnum\value{lstnumber}=11\color{BurntOrange}\fi%
\ifnum\value{lstnumber}=12\color{Turquoise}\fi%
}]
for (int i = str.Length - 1; i >= 0; i--) {
    char c = str [i];
    if (c == '\n') {
        if (index == -1)
            index = i;
        index = i;
    }
    else if (c == ' ' || c == '\t')
        continue;
    else
        break;
}
\end{lstlisting}
\caption{Example of code clone with nested syntax}
\label{lst:nested-for}
\end{figure}

\begin{table}[b]
\centering
\caption{Evaluation Results with Behavior Parameterization}
\label{tab:behaviourResults}
\begin{tabular}{lr}\hline
Category                  & Code clones \\ \hline
Refactored successfully   & 175         \\
Refactored, not tested    & 34          \\\hline 
Total                     & 209         \\\hline 
\end{tabular}
\end{table}

\subsection{Why does refactoring fail?}
Table~\ref{tab:results} shows that refactoring failed in 39.0\% of the refactorable code clones.
One factor contributing to refactoring failure is the use of private classes.
The class that defines the subroutine is external to the original classes that contain the code clones.
Therefore, it cannot access private classes used in the original source code.
The proposed approach does not take into account the accessibility of classes used within the code fragments.

To solve this problem, it is effective to check the accessibility of the used classes and members.
Another solution is to consider changing the location where the subroutine is defined.
Defining the subroutine in the class that contains the code clone can resolve accessibility issues and avoid problems related to instance method calls.
However, this approach cannot be used when the code fragments in the clone pair belong to different classes.

Additionally, relying on interfaces can sometimes resolve accessibility problems. 
If there is a public interface with a private implementation, using the interface instead of the implementation can avoid accessibility issues. 
This measure is effective if using the interface sufficiently replicates the original behavior.
In our variable mapping, our approach selected the most specific common class that closely matches the actual variable types.
To address accessibility issues, the parameter type should be selected as the most specific common class that is publicly accessible and closest to the actual variable types.

\subsection{How effective is behavior parameterization?}
The proposed approach performs behavior parameterization only when the clone pair contains statements that differ.
We investigated how many clone pairs were refactored using behavior parameterization.
Table~\ref{tab:behaviourResults} shows how many clone pairs were refactored using behavior parameterization.
Among the refactorable clone pairs, 27.0\% were successfully refactored using behavior parameterization. 
This result demonstrates that behavior parameterization is effective in a substantial number of cases.

\section{THREATS TO VALIDITY}
There are two primary threats to the validity of this study.

Internal validity may be threatened by the evaluation method.
To verify the correctness of refactoring, we compared test results. 
The quality of the tests affects the reliability of the results.
Since test results are not always stable, there is a risk of coincidental matches or mismatches.
To eliminate unstable tests, we ran the tests multiple times.
Additionally, to avoid false negatives from tests that do not cover the refactored portions, we introduced an incorrect subroutine generator.
We then evaluated whether the tests could detect the incorrect behavior.

External validity may be threatened by the quality of the evaluation data.
The results of the evaluation are affected by the quality of the code clone pairs used for testing.
To reduce bias in clone pair quality, we conducted evaluations across 22 different projects.
Furthermore, the quality of the clone pairs also depends on the code clone detector used.
In this study, we used NiCad~\cite{NiCad}.
Other code clone detectors capable of detecting clones in C\# include CCFinder~\cite{ccfinder} and Simian~\cite{simian}.
Future studies are encouraged to consider using different code clone detectors or conducting validation on different projects.

\section{CONCLUSION}
In this study, we proposed and implemented an Extract Method refactoring approach for C\#, which utilizes behavior parameterization.
Furthermore, we evaluated our approach using projects that included test cases.
We applied the proposed refactoring and analysis approach to 2,217 clone pairs detected from 22 C\# projects.
We then verified the correctness of the refactoring for the clone pairs that were successfully refactored.

The proposed approach determined that 35.0\% of all clone pairs were suitable for refactoring.

Furthermore, 61.0\% of the refactorable projects built successfully.
For 32.1\% of these projects, the provided tests passed, confirming the behavioral correctness.
In the remaining 28.9\% of cases, behavioral correctness could not be confirmed because the test suite did not cover the refactored source code.

In this study, we used the detection results from NiCad~\cite{NiCad}.
Future work could investigate how the results differ when using other code clone detectors.
Variable names in both the subroutine and the caller were generated mechanically using sequential numbering.
Improving readability through variable name refactoring~\cite{naming} is a promising direction for future work.

\begin{acks}
This work was supported by JSPS KAKENHI Grant Number 25K03102, 24H00692, 23K24823.
\end{acks}

\bibliographystyle{ACM-Reference-Format}
\bibliography{reference}

\end{document}